\documentclass[fleqn,twoside]{article}
\usepackage{gc}
\usepackage{amsfonts}
\heads{S. Cotsakis, P.G.L. Leach and H. Pantazi}
       {Symmetries of Homogeneous Cosmologies}
\def\dsp{\displaystyle}
\def\d{\mbox{\rm d}}

\def\ha{\tst\frac{1}{2}}
\def\tha{\tst\frac{3}{2}}

\def\p{\dsp\partial}
\def\arcsinh{{\rm arcsinh}}
\def\arccosh{{\rm arccosh}}

\def\ga{\alpha}
\def\gb{\beta}

\def\gg{\gamma}

\def\gl{\lambda}

\def\gG{\Gamma}
\def\gD{\Delta}

\def\gP{\Psi}

\def\p{\dsp \partial}

\def\de{{\d\over\d e}}

\def\dddot#1{\mathinner{\buildrel\vbox{\kern5pt\hbox{...}}\over{#1}}}

\def\be{\begin{equation}}
\def\ee{\end{equation}}
\def\bq{\begin{eqnarray}}

\def\abstract#1{\begin{minipage}{350pt}{\large\bf Abstract}: #1 \end{minipage}}

\def\de{differential equation}
\def\des{differential equations}

\def\pde{partial differential equation}
\def\pdes{partial differential equations}

\def\odes{ordinary differential equations}

\def\ie{{\it ie }}

\def\etal{{\it et al. }}
\def\n{\nonumber}
\def\({\left (}
\def\){\right )}
\def\bi{\begin{itemize}}
\def\ei{\end{itemize}}

\def\lb{\left[ }
\def\rb{\right] }

\def\lrl{Laplace-Runge-Lenz}

\def\re#1{(\ref{#1})}

\def\R{{\mathbb R}}
\def\tall {\mbox{$\vphantom{\raisebox{2.5mm}{0}}$}}
\def\Tall {\mbox{$\vphantom{\raisebox{5mm}{0}}$}}
\def\low {\mbox{$\vphantom{\raisebox{-5mm}{0}}$}}

\def\lr{\longrightarrow}
\def\etc{{\it etc.}}
\def\cf{{\it cf. }}

\bls{0.99}
\begin{document}
\thispagestyle{empty}
\twocolumn[

\Title
{SYMMETRIES OF HOMOGENEOUS COSMOLOGIES}

\Author{S. Cotsakis\foom 1, P.G.L. Leach\foom2 and H. Pantazi\foom 3}
          {Laboratory for Geometry, Dynamical Systems and Cosmology (GEODYSYC)\\
          Department of Mathematics, University of the Aegean,
          Karlovassi 83 200, Greece}

\Abstract
{We reformulate the dynamics of homogeneous cosmologies with a scalar
field matter source with an arbitrary self-interaction potential
in the language of jet bundles and extensions of vector fields. In this
framework, the Bianchi---scalar field equations become subsets of the
second Bianchi jet bundle, $J^2$, and every  Bianchi cosmology is
naturally extended to live on a variety of $J^2$. We are interested in
the existence and behaviour of extensions of arbitrary Bianchi-Lie and
variational vector fields acting on the Bianchi variety and accordingly
we classify all such vector fields corresponding to both Bianchi classes
$A$ and $B$. We give examples of functions defined on Bianchi jet bundles
which are constant along some Bianchi models (first integrals) and
use these to find particular solutions in the Bianchi total space.  We
discuss how our approach could be used to shed new light to questions like
isotropization and the nature of singularities of homogeneous cosmologies
 by  examining the behaviour of  the variational vector fields and also
give rise to interesting questions about the `evolution' and  nature of
the cosmological symmetries themselves.
}
]
\email 1 {skot@aegean.gr}
\email 2 {leach@math.aegean.gr}
\email 3 {hpant@aegean.gr}

\section{Introduction}

An Equivalence Problem asks whether two geometric objects (e.g., manifolds,
metrics, differential equations, Lagrangians, cosmological models etc) are
the same under a suitable change of variables. Symmetries of a geometric
object are defined as {\em self-equivalences} of the object and the
determination of the symmetry {\em group} of an object is a special case
of the general equivalence problem.  Two equivalent geometric objects have
isomorphic symmetry groups and, indeed, symmetry plays a central role in
equivalence since if the symmetry groups of two objects are not isomorphic
(e.g., of different dimensionality), the objects cannot be equivalent.

There are two main approaches to Symmetry Theory or Equivalence
Problems, that of Sophus Lie and that of Ellie Cartan. Lie's approach
originally attempted to classify all possible Lie groups of
transformations on one- or two-dimensional manifolds and
has recently attracted a lot of attention \cite{o1}, especially the newly
significant role it plays in problems of differential equations and
variational calculus.  On the other hand, the {\em coframe equivalence
problem}, which includes all other equivalence problems as particular
cases, introduced and algorithmically solved by Cartan, although it
initially triggered a lot of activity (mainly due to the effort of
Cartan's students), subsequently declined in applications due to its
calculational complexity. However, very recently it received renewed
attention and was revitalized primarily by Olver (see \cite{o2} and
references therein).  Lie's approach is more ``analytic'' than Cartan's,
which is more geometric in nature and is implemented by using the full
machinery of differential forms.

Equivalence problems in gravitation are abundant. For example, the conformal
equivalence \cite{baco} between different higher order theories of
gravitation and general relativity with an additional field is achieved via
a Legendre transformation which is a particular case of contact symmetry.
Usually the first step to solving an equivalence problem is the
determination of symmetries of the object in question. Since Einstein
equations are the Euler-Lagrange equations of the Hilbert action functional,
many equivalent problems in relativity are bound to involve {\em
variational\/} (Noether) symmetries, that is, point or generalized
transformations which leave the action functional invariant. Since every
variational symmetry of a (variational) problem is a (Lie) symmetry of the
associated Euler-Lagrange equations, one may first determine the complete
symmetry group of the equations and then decide which of these are
variational.

A different kind of equivalence problem appearing in relativity is that of
classifying the symmetries of particular relativistic {\em models}.
Although there is to date no systematic attempt to tackle such equivalence
problems, some interesting work has been done to determine the  symmetries
of some of the Bianchi cosmological models.  In particular, Capozzielo \etal
\cite{italians} classified the variational symmetries of some of the Bianchi
Lagrangians and gave (through the Noether theorem) some first integrals (in
particular, those for the Bianchi types I and V).

Since the Lie approach is more `easily' implemented and is less geometric in
nature, in this paper we focus on the implementation of this method to the
problem of finding the Lie symmetry groups of all Bianchi cosmologies in
vacuum or with a scalar field possessing an arbitrary self-interaction
potential.  We also give, when possible, the associated first integrals,
thus complementing the previous work \cite{italians}.

The plan of this paper is as follows. \sect 2 gives a brief
introduction to the basic ideas of the geometric theory of symmetries which
is fundamental to our work. In \sect 3 we apply the theory of \sect 2
to homogeneous cosmologies and, in particular, recast the Bianchi dynamics
in the language of jet bundles and extensions of vector fields and give
a full list of symmetries for the Bianchi-scalar models with an arbitrary
self-interaction potential. The use of these results is exemplified in
\sect 4 where we give explicit forms of first integrals and construct
solutions in many new cases of Bianchi-scalar  cosmologies. We
conclude in \sect 5 by pointing out several possible directions
for future research.

\section{Jet bundles, extensions and variational symmetries}

Although in this paper we focus on sets
of \odes, for the purpose of this section it is very convenient to
consider a general set of partial differential equations involving $p$
independent variables $x=(x^1,\ldots,x^p)$ and $q$ dependent variables
$u=(u^1,\ldots,u^q)$. The {\em total space} is the Euclidean
space\footnote
  {In the language of vector bundles $E$ is replaced by a vector
   bundle over the base space $M$ (see \cite{o1}) and $M$ and $N$ are
   differential manifolds, not necessarily submanifolds of Euclidean space.}
$E=M\times N=\{(x,u):x\in M\subset \R^p,\ u\in N\subset \R^q\}\subset
\R^{p+q}$. We restrict attention only to the most elementary type of symmetry
that is, {\em point symmetries} which are defined to be local
diffeomorphisms of $E$ onto itself, $g:E\longrightarrow E:(x,u)\mapsto
g(x,u)$
\be
    (\bar x, \bar u)=g(x, u)=(X(x, u), \gP(x, u))    \label{trf}
\ee
which act pointwise on $E$. The set of all such diffeomorphisms forms a group,
the {\em symmetry group\/} of the total space $E$.
A basic example of a point transformation is
constructed by  starting with a vector field on $E$
(that is a section of $TE$)
\bear
    v \eql \xi^i(x, u)\p_{x^i} +\phi^{\ga}(x, u)\p_{u^\ga},\nn
    && \qquad i=1,\ldots ,p,\cm \ga=1,\ldots ,q      \label{point}
\ear
and considering its flow $\exp (tv)$. This defines  a 1-parameter group of
point transformations of the total space $E$ onto itself. (In what follows,
in a slight abuse of language, we refer to a symmetry either as  $g$ or as
 its infinitesimal generator $v$.)

A function $f:M\lr N, \ u=f(x)$, (\ie a section of the bundle $E$) is {\em
invariant\/} under a group of transformations  $G$ if its graph $\gG _f=\{(x,
f(x)): x\in M \}\subset E$ is a {\em $G$-invariant\/} subset, which means
that $\forall g\in G$ and $(x, u)\in \gG _f$, $g(x, u)\in \gG_f$.  $\gG_f$
is a regular $p$-dimensional submanifold of $E$ although only transverse
submanifolds of $E$ give (locally)
graphs of functions of the total space (see \cite{o2}, p.\,107).
We must distinguish between an invariant function $u=f(x)$
and {\em an invariant\/} of $G$: A function $I(x,u)$ on $E$
\footnote{Notice that for this definition $x$ and $u$ are not necessarily
dependent variables.} such that $v(I)=0$. This is a direct generalisation,
on functions defined on $E$, of the usual concept of {\em an invariant\/} on
$M$, $f(x)$, which means a real-valued function $f\in C^{\infty}(M)$ such
that $v(f)=0$ where $v=\xi^i(x) \p_{x^i}$. (In this case we have a first
order, linear, homogeneous partial \de\ $v[u]=0$ which is solved by the
method of characteristics.) In all these cases $G$ is the {\em symmetry
group} and members of $G$ are called {\em the symmetries of the function}.

How do we characterize invariant functions? Since the graph $\gG_f$ of an
invariant function $u=f(x)$ is determined by the vanishing of its components
$u^\ga -f^{\ga}(x)$, $\ga=1,\ldots ,q$, a direct application of the {\em
infinitesimal invariance criterion} (\cf\ \cite{o2}, p.\,65) gives, using
(\ref{point}),
\bear
0 \eql v\lb u^{\ga}-f^{\ga}(x)\rb=\phi^{\ga}(x, u) -\xi^i(x, u)
                    \frac{\p f^{\ga}}{\p x^i}, \nnn
    \cm  \ga = 1,\ldots,q  \label{criterion}
\ear
for every infinitesimal generator $v$ of the form (\ref{point}). We call
the $q$-tuple in the right hand sides of (\ref{criterion}) {\em the
characteristic\/} of the vector field (\ref{point}) and denote it by
$Q^{\ga}(x, f^{(1)})$. Therefore
(\ref{criterion}) states that a function $u=f(x)$ defined
on the total space $E$ is
invariant under a (connected) group of point transformations if and only
if it satisfies the equations
\be
    Q^{\ga}(x, f^{(1)})=0, \cm  \ga=1,\ldots,q.
\ee

Since in what follows we shall be dealing with symmetries of differential
equations and of variational problems,  we need to have a definition of
symmetry for  functions that depend not only on the  $x$'s and $u$'s
but also {\em on the derivatives} of the dependent variables $u^{\ga}$ and
the independent variables $x^i$. This necessarily takes us away from the
total space $E$ with the coordinates $(x, u)$ and into the
higher-dimensional analogues of $E$ called jet bundles. This will also
require an extension of the infinitesimal invariance criterion
(\ref{criterion}) to deal with such functions.

Therefore we need to develop some aspects (particularly interesting from
the point of view of the applications in the next Sections) of the theory of
{\em jet bundles\/} and {\em extension of vector fields}.  Furthermore,
we shall have to recast \des\ in a more geometric language using the notion
of a {\em variety on the jet bundle}. Below, we give a very rapid overview
of only those elements of the relevant theory we need, referring to
\cite{o1} and \cite{o2} for a more complete discussion including references.

The notion of a jet bundle is very simple. We may think of it as a space
whose coordinates are the $p$ independent variables $x^i$, the $q$ dependent
variables $u^{\ga}$ and the derivatives of $u^{\ga}$, $\ga=1,\ldots,q$, of
orders 1 up to and including $n$. This is the {\em $n$-th jet bundle} of the
total space $E=M\times N\subset \R^p \times \R^q$ which we denote by
$J^n=J^nE=M\times N^{(n)}$ where
$M$ is the (base) space of the independent variables and $N^{(n)}$ is the
fiber containing the remaining variables (dependent plus their derivatives
of orders 1 to $n$). (A more rigorous way is to define $J^n$ as the set of
equivalence classes of $C^{\infty}$ functions wherein two functions are
equivalent  at $x$ if and only if they are in {\em $n$-th order contact at
$x$}, meaning that their Taylor polynomials at $x$ of order $n$ are
identical.)

Having `extended' the total space $E$ to the jet bundle $J^nE$, it is natural
to define the {\em $n$-th extension, $f^{(n)}$,
of a function $f:M\lr N$} to be a section
of $J^n$, \ie, $f^{(n)}:M\lr N^{(n)}$ is defined by evaluating all partial
derivatives of $f$ of order 1 to $n$. The graph of the extended function
$f^{(n)}$, $\gG _{f^{(n)}}=\{(x, f^{(n)}(x))\}$, will similarly be a
$p$-dimensional submanifold of $J^n$. Since the point transformation
(\ref{trf}) will also act on the derivatives of functions $f:M\lr N$, we can
define the {\em induced extended (point) transformation
$g^{(n)}:J^n \lr J^n$} by
\be
    \(\bar x, {\bar u}^{(n)}\)=g^{(n)}\(x, u^{(n)} \)    
\ee
on the $n$-th jet space. This will transform the graphs of {\em extended}
functions giving $g^{(n)}\gG _{f^{(n)}}=\gG _{g(f)^{(n)}}$ (\cite{o2},
p.\,113).

A smooth real-valued function $F:J^n \lr \R$ (or on an open subset of $J^n$)
is called a {\em differential function of order $n$}. An {\em $n$-th order
differential equation} is defined by the vanishing of an $n$-th order
differential function. The {\em total derivative}, $D_{x^i}F$, of a
differential function of order $n$ with respect to $x^i$, is an  $(n+1)$-th
order differential function defined in the usual way.  For instance, in the
case of one independent variable $x$ and one dependent variable $u$, we have
the following formula for the total derivative of $F(x, u^{(n)})$ with
respect to  $x$:
\be
    D_xF=F_x+u_xF_u+u_{xx}Fu_x+u_{xxx}Fu_{xx}+\ldots .
\ee
Obviously vector fields on $E$ can also be extended to vector fields on
$J^nE$.  The {\em $n$-th extension of a vector field $v$} with the
characteristic $Q=(Q^1,\ldots,Q^q)$ has the following form (see \cite{o1}
for a proof):
\be
v^{(n)}=\xi ^{i}(x, u)\p _{x^i}+{\phi _J} ^{\ga} (x, u^{(j)})
\p _{{u_J}^{\ga}},
\ee
where ${\phi _J}^{\ga} =D_JQ^{\ga}+{\xi}^iu^{\ga}_{J,i}$, with $i=1,\ldots,p$,
$\ga=1,\ldots,q$,  $|J|=j=0,\ldots,n$ ($J$ is an obvious  multi-index) and
summation over repeated indices is implied in the usual way.

A point symmetry of a system of (partial) differential equations is a point
transformation $g:E\lr E$ with the property that, if $u=f(x)$, is a solution,
then the transformed function $\bar u={\bar f}({\bar x})$ is also a
solution. Suppose now that we are given a family of differential functions
$\gD _{\gb}:J^n \lr R:(x, u^{(n)}) \mapsto \gD_{\gb}(x, u^{(n)})$, indexed
by $\gb=1,\ldots,m.$ A set of \des\ of order $n$ is defined by the
simultaneous vanishing of a given family of differential functions:
\be
        \gD _{\gb} (x, u^{(n)})=0.        \label{des}
\ee
It is very important to view these equations as defining, or defined by, a
{\em variety}
\be             \nhq
S_{\gD}=\{(x, u^{(n)}):\ \gD _{\gb}(x, u^{(n)})=0, \ \gb=1,\ldots,m\} ,
\ee
which is a subset of $J^n$ consisting of all points of $J^n$ which
simultaneously satisfy \eqs (\ref{des}). Thus in our geometric
reformulation a set of differential equations is a {\em (sub)set
of some space}.  A {\em solution} of \eqs (\ref{des}) is a function $u=f(x)$
on $E$ such that the graph of its $n^{th}$ extension, $\gG _{f^{(n)}}$,
lies entirely on $S_{\gD}$
(this is translated traditionally as  `the function $u=f(x)$ identically
satisfies \eqs (\ref{des})'.

It follows that the defining property of a symmetry of  a differential
equation discussed above, as a transformation that maps solutions into
solutions, can be geometrically reformulated in an elegant manner by simply
requiring that
\be
g^{(n)}(S_{\gD})\subset S_{\gD}.
\ee
On the other hand, the infinitesimal invariance criterion, (3),
immediately implies that
the fundamental condition for $v$ to be a symmetry of the fully
regular\footnote{This, roughly speaking, means that the system {\em has}
solutions} system of \des\ (\cite{o2}, p. 179) is
\be
v^{(n)}(\gD _{\gb})=0.  \label{condition}
\ee
Notice that  $v^{n}$ acts only on those points on $S_{\gD}$ which lie on a
solution.  \eq (\ref{condition}) typically results in a large,
overdetermined, linear system of \pdes\ and its solution requires very
lengthy calculations which are usually performed using some of the computer
algebra packages available (see below, also \cite{wym}).

In (\ref{point})
$\xi ^i$ and $\phi^{\ga}$ depend only on the total
space coordinates $x$ and $u$, and this was assumed by Lie \cite{lie}.
Subsequently, he extended that dependence to include first or higher order
derivatives of the functions $u^{\ga}$, $\ga=1,\ldots,q$. Such generalized
vector fields produce the so called {\em contact\/} and {\em generalized\/}
symmetries. More recently, {\em nonlocal symmetries\/} \cite{nl1, nl2},
containing {\em integrals\/} of the dependent variable(s), have found their
use.  Thus there is a hierarchical sequence of symmetries: point, contact,
generalized, nonlocal.

Once the symmetries of a differential equation have been determined, one may
effectively reduce the order of the equation by calculating first
integrals. For a set of differential equations, $\gD_{\gb}(x,u^{(n)})=0$,
{\em a first integral\/} is a function $I(x,u^{(m)} )$  which is
constant along solutions, that is
\be
        D_{x}I=0.                         
\ee
This equation is typically a linear \pde\ and the associated characteristic
system provides a means to calculate all first integrals. If all of them can
be determined, (8) is said to be integrated.

The complete solution of the (nontrivial) characteristic system of (18) is
rarely trivial and one seeks to ease the process of solution by imposing the
condition that any first integral be associated with a symmetry of (8),
i.e., we require that
\be
    v^{(n)} (I)=0,
\ee
in addition to (8). This equation is also a linear \pde\ for I and so the
number of characteristics is further reduced. Usually, however, the
existence of suitable symmetries (not necessarily point ones) equal in
number to the order of the system is required, but the cost is the reduction
of one first integral for each requirement of invariance under symmetry
imposed.

The use of Lie symmetries to determine first integrals can be extremely
difficult since the calculations to be performed {\em after} the symmetries are
obtained are usually difficult. The fundamental theorem of Noether \cite{noe}
provides an attractive alternative. A point (or generalized)
transformation is called a {\em variational (or Noether)
symmetry} of the functional
\be
    A[u]=\int_{\Omega}L(x,u^{(n)})\d x ,\hspace{1cm}\Omega\subset M
\ee
if and only if the transformed functional agrees with the original one. The set of all
variational symmetries forms a group and this  {\em variational symmetry
group\/} is a symmetry group of the associated Euler-Lagrange equations but
not conversely (cf. \cite{o2}, p.\,236). The infinitesimal invariance
criterion applied to a (connected) transformation group $G$ gives the basic
condition for $G$ to be a variational symmetry group of $A[u]$, namely,
\be
    v^{(n)}(L)+L\, {\rm div}\;\xi =0,
\ee
for every infinitesimal generator $v$, where $L$ is the Lagrangian of the
variational problem and $\xi =(\xi^i )$ is the vector of basic
components in (2) (\cf\ \cite{o2}, p.\,236).

As in the case of the Lie symmetries of \des, the variational symmetries and
consequent variational first integrals impose no requirements on the
coefficient functions $\xi$ and $\phi$ apart from differentiability.
Variational symmetries can be point, generalized (including contact) or
nonlocal.  The latter are not of practical use for a local Lagrangian unless
the nonlocal terms cancel in such a way that the first integral is local.

Given the ease of computation of variational integrals once the variational
symmetries are known, one may wonder at the interest in the more difficult
calculation of first integrals using the Lie symmetries of the \de.  For
instance, provided
the Hessian of the Lagrangian with respect to the first derivatives
 is nonsingular, the Euler-Lagrange system is $2n$-dimensional
and gives a regular Hamiltonian system.  Each variational symmetry reduces
the dimension of the system by two.  If there exist $n$ independent
variational integrals, there are $n$ integrals in involution and the system
is integrable according to Liouville's Theorem.

Consequently, variational symmetries are very attractive.  However, the
calculation of variational symmetries is a closed procedure only in the
case of variational {\em point} symmetries.  In the case of
generalized symmetries some {\it ansatz\/} must be made and there is
always a possibility that some symmetry and so its integral will be missed.
(There are cases for which variational symmetries are known to be
generalized and yet the integral corresponds to a Lie point symmetry.
Perhaps the best-known of these are the \lrl\ components which
correspond to generalized variational symmetries linear in the derivatives
whereas the corresponding Lie symmetry is point \cite{p81}.)  The fear of
incompleteness in the knowledge of variational symmetries in complex
systems is a sufficient incentive to supplement the variational method with
the Lie method.

In this paper, we restrict ourselves to point symmetries in both the
variational and Lie approaches. (The calculation of other types
of symmetry is extremely difficult even for scalar ordinary
differential equations and often does not give more information
than obtained from the point symmetries.) We have remarked that
the variational integrals follow easily once the variational
symmetries are known.  We use the Lie method to supplement the
variational approach.  Indeed, there are some instances in which
we can combine the two methods in the case that a variational
integral is invariant under a Lie symmetry, which may or may not
be the variational symmetry of the integral.
The knowledge of this integral, both Lie and variational, can help one
in solving the associated Lagrange system for the other integrals.

Our strategy is now clear and will be described in detail in the next
sections.  We start by defining the Bianchi total space $E_B$ to be  the
space including time plus the dependent variables, extend $E_B$ to the
second Bianchi jet bundle $J^2_B$ and consider extensions of
all basic functions defined on the Bianchi total space (such as the Ricci
scalar, \etc) to $J^2_B$. In this way the field equations become conditions
on $J^2_B$ defining a Bianchi variety, $S_{\gD ,B}$. This is a  subset of
the second Bianchi jet bundle containing all Bianchi models and consequently
the structure of $S_{\gD ,B}$ is of fundamental importance to our work.  The
point symmetries analysed in the next section  have the property that, when
extended, they leave $S_{\gD ,B}$ invariant. (This property is also shared
by any contact vector fields although in this case (which do not consider in
this paper) the Bianchi Lagrangian forms will not be invariants of the
Bianchi jet bundle and will have to be modified to include Cartan extensions
known as coframes (\cf \cite{o2}).)

\section{Point symmetries of homogeneous cosmologies}

In Bianchi homogeneous but anisotropic models the spacetime
metric splits so that the spatial (time--dependent) part is given by
\[
  g_{ab}(t)=\exp (2\lambda )\exp (-2\beta )_{ab}
\]
where $\lambda$ plays the role of a time (volume) parameter and
$\beta$ is a $3\times3$ symmetric, traceless matrix which can be
written in a diagonal form with two independent components by
introducing the two {\em anisotropy parameters} $\beta _{+},\ \beta _{-}$:
\[
\beta _{ab}=\diag\biggl(\beta _{+},
    -{1\over 2}\beta _{+}+{\sqrt {3}\over 2}\beta
       _{-},-{1\over 2}\beta _{+}-{\sqrt {3}\over 2}\beta_{-}\biggr).
\]

The general Lagrangian leading to the full Bianchi--scalar dynamics has the
form (see e.g. \cite{maccallum})
\bearr                                                    
  {\cal L}= \e^{3\gl}\biggl[ R^ {*}
      +6{\dot \gl} ^2- \tha \({\dot {\gb}_1} ^2+ {\dot {\gb}_2} ^2 \)
           -{\dot \phi} ^2+2V\(\phi\) \biggr],     \nnn       \label{hLagr}
\ear
where $R^*$ is the Ricci scalar playing the role of a potential term. The
{\em Bianchi\/} total space is $E_B=\{ (t,\lambda , \beta_{+} ,
\beta_{-} ,\phi )\}\subset \R^5$. Then the first and second
Bianchi jet bundles are $J_B^{1}=E_{B}\times\{ (\dot{\gl},\dot{
\beta_{+}},\dot{\beta_2},\dot{\phi})\}\subset \R^9$ and
$J_B^{2}=J^1_B\times\{ (\ddot{\gl},\ddot{ \beta_{+}},\ddot{
\beta_{-}},\ddot{\phi})\}\subset \R^{13}$.

The Euler-Lagrange equations for (\ref{hLagr}) can be considered as
\eqs (\ref{des}) with $m=4$ and $n=2$. Explicitly (\ref{des}) becomes
\bear
\ddot {\gl}+ \frac{3}{2} {{\dot \gl} ^2}+\frac{3}{8}\({\dot {\gb}_1} ^2 +
    {\dot {\gb}_2} ^2 \) +\frac{1}{4} {\dot \phi}^2 \cm    \lal\nn
 - \frac{1}{12} \e^ {-3\gl}
        {\p \over \p{\gl}}\(\e^{3\gl} R^ {*}\)
                    -\frac{1}{2} V\(\phi\)\eql 0, \nn
 {\ddot {\gb}_1}  +  3{\dot {\gb}_1}{\dot \gl}+\frac{1}{3}
                              {\p{R^ {*}} \over \p{\gb_1}} \eql 0 \nn
                                                 \label{hsys}
 {\ddot {\gb}_2} + 3{\dot {\gb}_2}{\dot \gl}+\frac{1}{3} {\p{R^{*}}
                                            \over \p{\gb_2}} \eql 0 \nnv
          {\ddot \phi} + 3{\dot \phi}{\dot \gl}+V' \eql 0.     
\ear
For Bianchi class A models the Ricci scalar $R^*$ as a function on a
four-dimensional hypersurface of the Bianchi
total space $E_B$ has the explicit form
\bear
 R^{*} \eql - \Half \e^ {-2\gl}\biggl[ {N_1} ^2 \e^ {4{\gb}_1}     \nnn
      +\e^{-2{\gb}_1}
              \(N_2 \e^ {{\sqrt 3} {\gb}_2}-N_3 \e^ {{-\sqrt 3}{\gb}_2}\)^2
                                 \nnn
      -  2{N_1}{\e^ {{\gb}_1}} \({N_2}\e^ {{\sqrt 3}{\gb_2}}
                    +{N_3}\e^ {{-\sqrt 3}{\gb_2}}\) \biggr]       \nnn
      \cm +\Half {N_1}{N_2}{N_3}\(1+ {N_1}{N_2}N_3 \) ,                
\ear
and for class B
\bear
    R^* =2a^2\e^{-2\gl}\(3-\frac{N_2 N_3}{a^2} \)\e^{\gb}
\ear
with
\be
    \gb=\frac{2}{3a^2-N_2 N_3} \(N_2 N_3 {\gb_1}
                        +{\sqrt {-3a^2 N_2 N_3} \gb_2} \),
\ee
where $a, N_1, N_2, N_3$ are the usual classification constants.
For the subsequent symmetry analysis we set
\be
u= \e^ {\gl}, \qquad v=\e^ {{\gb}_1}, \qquad w = \e^ {{\sqrt 3}{\gb_2}}.
\ee
In what follows
we list the relevant symmetries of  different Bianchi models in
four cases according to different matter couplings:
\begin{description}
\item
[Case 1: Vacuum.] It is obvious that the potential and $\phi$ terms
are missing in the $u$ equation and also there is no $\phi$ equation.  Hence
we have a set of three equations with three dependent  variables $u$, $v$
and $w$. This is the easiest case of all the models and, once having
examined it, one is able to make inferences about the forms of the
symmetries in the subsequent cases.
\item
[Case 2: Scalar field with zero potential.] In this case we have a set of
four equations with four dependent variables $u,v,w$ and $\phi$. The
 potential term is missing in the $u$ and $\phi$ equations.
\item
[Case 3: Constant potential.] The equations have the same form as in
Case 2 plus a constant potential term in the $u$ equation.
\item
[Case 4: Arbitrary potential.] It is the most general case.
We have a set of four equations with four dependent
variables. Moreover, in the $u$ equation there is a
potential function and in the $\phi$ equation we have a derivative of this
potential with respect to $\phi$.
\end{description}
As we shall see, extra symmetries arise in cases 2--4. This is
because the first three equations in (23) have $\phi$ as an
ignorable coordinate in cases 2 and 3 and so this is a symmetry
of the fourth equation in (23). In case 1, $\phi$ is not an
argument of the coefficient functions.

We also note that the actual calculation of Lie symmetries in
this paper use, to a certain extent, the package {\sc LIE} (cf.
\cite{head}) which has been around for twenty years. The
equation and type of symmetry sought are fed into an input file.
The program computes the determining equations and then
attempts to solve them. Usually, however, this proves impossible
and in this case the operator can intervene manually.

\subsubsection*{Bianchi Type I}

This is the easiest case since the Ricci scalar is zero.

\medskip\noi
{\bf Case 1:} The Noether symmetries are
\[
   \p_t, \qquad  v\p_v, \qquad w\p_w,  \qquad v\log w\p _v-3w\log v\p _w.
\]
The additional Lie symmetries are
$   t\p _t, \ ut\p_u+\tha t^2\p _t$  and  $u\p _u$.

\medskip\noi
{\bf Case 2:} The extra variable $\phi$ gives three additional symmetries
which are also Noether symmetries:
\bearr
\p _t, \quad v\p _v, \quad w\p _w,
\quad {\p _{\phi}}, \quad v\log w\ \p_v-3w\log v\ \p_w,   \nnn
\cm v\phi{\p_v}-\tha  \log v{\p_{\phi}},
    \qquad w\phi\p _w-\half \log w{\p _{\phi}}.
\earn
The additional Lie symmetries are
\[
t\p _t, \qquad ut\p _u+\tha  t^2\p _t \  {\rm \ and} \ \ u\p _u.
\]

\medskip\noi
{\bf Case 3:} The extra term of the constant potential $V(\phi )=C$ affects
 the results and we obtain the following Noether symmetries:
\bearr
\p _t, \quad v\p _v, \quad w\p _w,\quad {\p _{\phi}}, \quad
                v\log w\p _v-3w\log v\p _w,          \nnnv
\cm  v\phi{\p _v}-\tha  \log v \p _{\phi}, \quad
                w\phi\p _w-\half  \log w{\p _{\phi}},
\earn
and the Lie symmetries
\bearr   \nq\,
u\p _u, \quad
\e^{\sqrt 3 Ct }{\p}_t+\e^{\sqrt 3 Ct} u {\p}_u,
\quad
\e^{\sqrt 3 Ct }{\p}_t-\e^{\sqrt 3 Ct} u {\p}_u.
\earn

\medskip\noi
{\bf Case 4:}
Since this is the most general case, the number of symmetries is obviously
reduced. The Noether symmetries are
\[
\p _t, \qquad v\p _v, \qquad w\p _w, \qquad v\log w\p _v-3w\log v\p _w,
\]
and the additional Lie symmetry is $u\p _u$.


\subsubsection*{Bianchi Type II}

The Ricci scalar takes the form $R^{*}=- \e^{(4\gb_{1}-2\gl)/2}.$

\medskip\noi
{\bf Case 1:}
The system has the Noether symmetries $\p _t$ and  $w\p _w$.
The additional Lie ones are $t\p _t+u\p _u$ and $-2t\p _t+v\p _v.$

\medskip\noi
{\bf Case 2:}
We obtain the additional Noether symmetry ${\p _{\phi}}$
and the additional Lie symmetry $w\phi\p _w$ $+\half \log w\p _{\phi}$.

\medskip\noi
{\bf Case 3:}
In this case the model has the Noether symmetries
$\p _t$, $w\p _w$ and ${\p _{\phi}},$
and the additional Lie symmetry $w\phi\p _w+{\half }\log w{\p _{\phi}}.$

\medskip\noi
{\bf Case 4:}
In this case we obtain $\p _t$ and  $w\p _w$ as Noether symmetries and
$u\p _u+\half v\p _v$ is a Lie symmetry.

\subsubsection*{Bianchi Type III}

The Ricci scalar is
$
R^{*}=8\e^{-2\gl}\e^{\frac{1}{2}\({\sqrt 3}\gb_2-\gb_1\)}.
$
The symmetry $\p _t$ is always a Noether one and $\p _{\phi}$ is
also  Noether for the cases where it exists.

\medskip\noi
{\bf Case 1:}
The Lie point symmetries are
\[
\p _t, \quad t\p _t+4v\p _v, \quad u\p _u+4w\p _w, \quad u\p _u-4v\p _v.
\]

\medskip\noi
{\bf Case 2:}
In this case we obtain the following Lie point symmetries:
\bearr
\p _t, \quad
 t\p _t+4v\p _v, \quad u\p _u+4w\p _w, \quad u\p _u-4v\p _v, \nnnv
\quad \p_{\phi}, \qquad v\phi\p _{v}+w\phi\p _{w}
              -\half \(3\log v+\log w\)\p _{\phi}.
\earn

\medskip\noi
{\bf Case 3:} The Lie point symmetries are
\[
\p _t, \qquad  u\p _u+4w\p _w, \qquad \p _{\phi},\qquad u\p _u-4v\p_v.
\]

\medskip\noi
{\bf Case 4:} Arbitrary potential.
We obtain the same symmetries as in the previous case apart from the
$\p _{\phi}$ symmetry, which is lost.


\subsubsection*{Bianchi Type V}

The Ricci scalar has the form $R^{*}=6\e^{-2\gl}.$

\medskip\noi
{\bf Case 1:} We obtain the Noether symmetries
$\p _t$, $v\p _v$ and $w\p _w.$ The additional Lie symmetries are
$t\p _t+u\p _u$ and $v\log w\p _v-3w\log v\p _w.$

\medskip\noi
{\bf Case 2:}
We find the additional Lie point symmetries
\[
2v\phi\p _v-3\log v{\p _{\phi}}, \qquad 2w\phi\p _w-\log w{\p _{\phi}}
\]
plus the Noether symmetry ${\p _\phi}.$

\medskip\noi
{\bf Case 3:}
We obtain the Noether symmetries
\[
    \p _t, \qquad v\p _v, \qquad w\p _w, \qquad {\p _{\phi}}.
\]
The additional Lie symmetries are
\bearr
v\log w\p _v-3w\log v\p _w,\cm v\phi\p _v-\fract{3}{2}\log v\p _{\phi},\nnnv
w\phi\p _w-\half \log w\p _{\phi}.\n
\ear

\medskip\noi
{\bf Case 4:}  The additional Lie point symmetries are
\[
    v\p _v, \qquad w\p _w, \qquad v\ln w\p _v-3w\ln v\p _w.
\]
There is only one Noether symmetry, $\p _t$.

\subsubsection*{Bianchi Type VI, class A}

The Ricci scalar has the form
\[
 R^{*}=-\ha \e^{-2\gl}\biggl[\e^{4\gb_1}+\e^{-2\(\gb_1-{\sqrt 3}\gb_{2}\)}
         +2\e^{\gb_1+{\sqrt 3}{\gb_{2}}}\biggr].
\]
The only Noether symmetries are $\p _t$ and
$\p _{\phi}$ in the cases where it exists.

\medskip\noi
{\bf Case 1:}
The Lie symmetries are
\[
  \p _t, \quad t\p _t-\half  v\p _v-\tha  w\p _w, \quad
{u\p _u+\half  v\p _v+\tha  w\p _w}.\n
\]

\medskip\noi
{\bf Case 2:}
There are the same Lie point symmetries as in the previous case.

\medskip\noi
{\bf Case 3:}
We find the following Lie point symmetries:
\[
    \p _t, \qquad u\p _u+\half  v\p _v+\tha  w\p _w, \qquad {\p _{\phi}}.
\]

\medskip\noi
{\bf Case 4:}
The system has the Lie symmetry $u\p _u+\half  v\p _v+\tha  w\p _w$.


\subsubsection*{Bianchi Type VI, class B}

The Ricci scalar takes the form
\[
R^{*}=2\(3a^2+1\)\e^{-2\gl}\exp{\lb{\frac{2}{3a^2+1}}
 \({\sqrt 3}a\gb_2-\gb_1\)\rb }. \n
\]
This model has only the standard Noether symmetries.
If we apply Program {\tt LIE} to this system,
a number of difficulties appear.  We set
\be
\omega = w ^{2a/(3a^2+1)}, \qquad n=v^{2/(3a^2+1)}
\ee
and further simplify  the system by defining the two constants
\be
B=\half (3a^2+1), \cm     C=1/(3a^2).
\ee

\medskip\noi
{\bf Case 1:}
In this case we obtain the following Lie point symmetries:
\bearr
\p _t,  \qquad t\p _t+u\p _u, \nnnv
-\half u\p _u+n\p _{n}, \qquad \half u\p _u+{\omega}\p _{\omega}.
\earn

\medskip\noi
{\bf Case 2:}
The Lie point symmetries are
\bearr
\p _t, \qquad t\p _t+u\p _u, \qquad -\half  u\p _u+n\p _n,\nnnv
{\half  u\p _u+\omega\p _{\omega}}, \qquad   \p _{\phi},\nnnv
 n\phi \p _{n}
        +\omega \phi \p_ {\omega}-\tha  B^2 C \log \omega \ \p _{\phi}
                    -\tha  B^2 \log n \ \p _{\phi}.
\earn

\medskip\noi
{\bf Case 3:} In this case we find the following results:
\bearr
   \p _t, \qquad -\half  u\p _u+n\p _{n}, \qquad
        \half u\p_u+\omega\p_{\omega}\p _{\phi},  \nnn
   n\phi \p _{n}+\omega \phi \p_{\omega}-
        \tha  B^2 C\log \omega \p _{\phi} -\tha  B^2 \log n \p _{\phi}.
\earn

\medskip\noi
{\bf Case 4:} We obtain the symmetries
\[
\p _t, \qquad \half  u\p _u+n\p _{n}, \qquad \half  u\p _u+{\omega\p _{\omega}}.
\]
The only Noether symmetries are the usual ones.

\subsubsection*{Bianchi Type VII}

In this model the Ricci scalar takes the form
$$
R^{*}=-\half  \e^{-2\gl}\(\e^{4\gb_1}+\e^{-2\(\gb_1-{\sqrt 3}\gb_{2}\)}
         -2\e^{\gb_1+{\sqrt 3}{\gb_{2}}}\).
$$

\medskip\noi
{\bf Case 1:}   There are three Lie point symmetries:
\bearr
\p _t, \qquad t\p _t-\half  v\p _v-\tha  w\p _w,
            \qquad u\p _u+\half  v\p _v+\tha  {w\p _w}.
\ear

\medskip\noi
{\bf Case 2:}
We have the additional symmetry ${\p _{\phi}}.$

\medskip\noi
{\bf Case 3:}
The symmetries of the system are
\[
\p _t, \qquad u\p _u+\half  v\p _v+\tha  w\p _w, \qquad {\p _{\phi}}.
\]

\medskip\noi
{\bf Case 4:}
We lose the ${\p _{\phi}}$ symmetry of the previous case.
In all cases the only Noether symmetries are the standard ones.

\subsubsection*{Bianchi Type VIII}

The Ricci scalar is
\bear
R^{*} \eql -\half  \e^{-2\gl}\biggl[ \e^{4\gb_1}
+\e^{-2\gb_1}\(\e^{{\sqrt 3}\gb_2}+\e^{{-\sqrt 3}\gb_2}\)^2\nnn
\cm
  -2\e^{\gb_1}\(\e^{{\sqrt 3}\gb_2}-\e^{{-\sqrt 3}\gb_2}x\) \biggr].
\earn

\medskip\noi
{\bf Case 1:}
The symmetries of the system are $\p _t$ and $t\p _t+u\p _u$.

\medskip\noi
{\bf Case 2:}
We obtain the additional symmetry ${\p _{\phi}}.$

\medskip\noi
{\bf Case 3:}
The system has only the two symmetries
$\p _t$ and ${\p _{\phi}}.$

\medskip\noi
{\bf Case 4:}
The only symmetry is $\p _t.$

\subsubsection*{Bianchi Type IX}

The Ricci scalar is
\bearr
 R^{*}= -\half  \e^{-2\gl}\biggl[ \e^{4\gb_1}
 +\e^{-2\gb_1}\(\e^{{\sqrt 3}\gb_2}-\e^{{-\sqrt 3}\gb_2}\)^2   \nnn
\cm\cm
    -2\e^{\gb_1}\(\e^{{\sqrt 3}\gb_2}+\e^{{-\sqrt 3}\gb_2}x\)\biggr]+1.
\earn
In Cases 1 and 4 the system has the single symmetry $\p _t.$
The other two cases have the additional symmetry ${\p _{\phi}}.$
\\

We now pass on to giving some examples of first integrals associated with the
symmetries obtained, postponing their discussion till Section V.

\section{Applications}

In this section we use the general theory developed in \sect 1 for
obtaining first integrals corresponding to symmetries for
Bianchi Types I, III and V and integrate to find some solutions.  It
is of interest that the expressions for the integrals are almost the same
for Cases 2 and 3.  We also describe how the existence of  an additional
variational symmetry restricts the possible forms the potential can take to
simple exponentials.

\subsection{Examples of first integrals}
\subsubsection{Bianchi Type I}

{\bf Case 1:} The results are given collectively in  Table 1 where
\begin{table}
\caption{First integrals for the Bianchi model type I in Case 1
         where there is no matter.}
\begin{center}
\begin{tabular}{|c|c|} \hline
Symmetry & Integral \\ \hline
        \tall
$v\p _v$ & \ \  $I_1=u^3 {\dot v}/{v} $ \\
        \tall
$w\p _w$ & \ \ $I_2=u^3 {\dot w}/{w} $ \\
$v \log w\p _v -3w\log v\p _w$
         & \low                 \Tall
      \  $I_3=\frac{1}{2}(\log u)^3 \fracd{\dot u{}^2}{u} -f_1(u)$ \\
      \hline
\end{tabular}                             \label{hiIcase1}
\end{center}
\end{table}
\bear
f_1(u) \eql \frac{1}{16}\(I_1 ^2 +I_2 ^2 \) \biggl[(\log u)^3\nnn
           \cm  +\frac{1}{2}(\log u)^2
                + \frac{1}{6} \log u+\frac{1}{36} \biggr] u^{-6}.
\earn
Using $I_3$, we find
\bear
    \dot u = {\sqrt  {g_1(u)}}
    \ \iff \ t-t_0 = \int \frac{\d u}{\sqrt{g_1(u)}}   \label{hint2}
\ear
where
\be
g_1(u)=\frac{2u^2 I_3}{(\log u)^3} + \frac{2u^2 f_1(u)}{(\log u)^3}.
\ee
We formally invert (\ref{hint2}) to find $u(t)$. The variables $v$ and
$w$ follow from the integrals $I_1$ and $I_2$, respectively. Hence we have
\[
    v =\exp {\dsp{\int \frac{I_1}{u^3} \d t}} ,  \cm
    w =\exp {\int {\frac{I_2}{u^3} \d t}}.
\]
We note that in this case we find the solution using only three
integrals since they are separable Noetherian integrals.

\medskip\noi
{\bf Case 2:} Setting
\bear
\ga \eql  \fracd{2}{3 \sqrt{K}}, \qquad
\gb = \lb \frac{2}{3} \frac{I_3 ^2}{K^2} \(K - 6I_4 ^2 \) \rb^{1/2}, \nn
 C \eql  \fracd{2 I_3 I_4}{K},  \qquad
 K =  \fracd{I_1 ^2 +I_2 ^2}{I_3 ^2},
\earn
we obtain the integrals listed in Table 2.
\begin{table}
\caption{First integrals for the Bianchi model type I in the case where there
is matter but no potential.}
\begin{center}
\begin{tabular}{|c|c|} \hline
Symmetry & Integral \\ \hline
    \tall
$v\p _v$ & \ \  $I_1=u^3 {\dot v}/{v} $ \\
    \tall
$ w\p _w$ & \ \ $I_2=u^3 {\dot w}/{w} $ \\
    \tall
$ \p _{\phi}$ & \ \ $I_3=u^3 {\dot \phi}$  \\
     \low  \Tall
$ v\log w\ \p _v$ & $I_4=\dsp{\frac{u^3}{I_3}\lb {\frac{\dot u}{u}}^2
                          -\frac{1}{4}I_1-\frac{1}{6}I_3 ^2 u^{-6} \rb}$ \\
    \low       \Tall
$ -3w\log v\ \p _w$
               & $I_5=t- \ga \arcsinh {\dsp{\frac{u^3 +C}{\gb}}}$   \\
       \hline
\end{tabular}
\label{hIIcase2}
\end{center}
\end{table}
(Note that $I_5$ may have $\arccosh$ instead of
$\arcsinh$, as we find after the calculation of the Lie integrals.)
Integrating, we obtain the following solutions:
\bear
u \eql  {\lb \gb\sinh (t-I_5)-C \rb} ^{1/3},   \nnv
v \eql  \exp \int {\frac{I_1}{u^3}} \d t               \nnv
 \eql \exp \biggl\{ I_1 \frac{1}{\sqrt{\gb^2+C^2}}\nnn
   \times\log \frac{\gb
     \tanh [(t-I_5)/2] -\gb+\sqrt {\gb^2+C^2}}
    {-C \tanh [(t-I_5)/2] -\gb-\sqrt{\gb^2+C^2}} \biggr\}, \nnv
w \eql
     \exp \int {\frac{I_2}{u^3}} \d t   \nnv
  \eql \exp \biggl\{ I_2 \frac{1}{\sqrt{\gb^2+C^2}} \nnn
       \times\log \frac{\gb
     \tanh [(t-I_5)/2] -\gb+\sqrt{\gb^2+C^2}}
    {-C \tanh [(t-I_5)/2] -\gb-\sqrt{\gb^2+C^2}} \biggr\}.
\earn

\medskip\noi
{\bf Case 3:}
The integrals have the same form as in the previous case, the only difference
being in the expression for  $\gb$ where an extra term due to the potential
is present,
\beq
\gb=\lb \frac{2}{3} \frac{I_3 ^2}{K^2}
           \(K - 6I_4 ^2 \) -\frac{C}{2K}\rb ^{1/2}.
\eeq


\subsubsection{Bianchi Type III}

The integrals in Cases 2 and 3 are essentially the
same. The integration becomes more involved in Case 3.

\medskip\noi
{\bf Case 1:} With the notations
\begin{eqnarray}
    \ga \eql-2\log u-{\frac{1}{2}}\log v+{\frac{1}{2}}\log w, \nn
    \gb \eql{\sqrt 3}\log v+{\frac{1}{\sqrt 3}}\log w,  \nn
    \gg \eql\log u,
\end{eqnarray}
the symmetry $v\p _v+w\p _w$  transforms to
${\p _{\gb}}$ with the associated Lagrange system
\be
\frac{\d t}{0}=\frac{\d \ga}{0}=\frac{\d \gb}{1}=\frac{\d \gg}{0}=\frac{\d
{\dot \ga}}{0}=\frac{\d {\dot \gb}}{0}=\frac{\d {\dot \gg}}{0}.
\ee
The characteristics of this system are
\bear
p \eql t, \qquad  u=\ga, \qquad w=\gg,\nn
x \eql \dot \ga, \qquad y=\dot \gb, \qquad  z=\dot \gg.
\ear
Using these characteristics, we obtain
\bearr
\frac{\d p}{1} = \frac{\d u}{x}=\frac{\d w}{z} =\frac{\d x}{\frac{3}{4}
\(x^2+y^2 \)-4 \e^u}\nnn
=\frac{\d y}{-3xy}
=\frac{\d z}{-\frac{3}{8}\(x^2+y^2+4xz+8z^2\)+\frac{2}{3}\e^u}
\earn
and thus
\bear
\frac{\d u}{x}  \eql \frac{\d y}{-3xy}
        \ \iff \ I_1 = \e^{3 \ga}{\dot \gb},\label{gob}    \yy
\frac{\d u}{x}
       \eql
      \frac{\d x}{\frac{3}{4}\(x^2+y^2 \)-4 \e^u} \nnn  \cm\cm
      \iff I_2 = \frac{{\dot \ga}^2}{2{\ga}^{3/2}} -f(\ga)\label{gab},\yy
\frac{\d p}{1}
       \eql  \frac{\d u}{x}
      \ \iff\ I_3 = t-\int\dsp{{\frac{\d \ga}{\dot \ga}}},
\ear
where
\be
f(\ga)={\frac{3}{4}}I_1 ^2 \int {\ga ^{-3/2}\e
^{-6\ga}\d\ga}+4\int{\ga^{-\frac{3}{2}}\e^{\ga}\d a}.
\ee
(Note that $\gb(t)$ follows from the quadrature of \re{gob}.)

The associated Lagrange's system for the symmetry $\p _t$  is
\be
\frac{\d t}{1}=\frac{\d u}{0}=\frac{\d v}{0}=\frac{\d \gg}{0}=
\frac{\d x}{0}=\frac{\d y}{0}=\frac{\d z}{0}
\ee
and  gives the characteristics
\bearr
    u=\ga, \qquad v=\gb, \qquad w=\gg, \nnn
     x=\dot \ga, \qquad y=\dot \gb, \qquad   z=\dot \gg.
\earn
From the system
\bearr
\frac{\d u}{x} = \frac{\d v}{y}=\frac{\d w}{z}=\frac{\d x}{\frac{3}{4}
            \(x^2+y^2 \)-4\e^u}\nnn
= \frac{\d y}{-3yx} =
     \frac{\d z}{-\frac{3}{8} \(x^2+y^2+4xz+8z^2\)+\frac{2}{3}\e^u}
\earn
we have
\bear
\frac{\d u}{x}\eql
\frac{\d z}{-\frac{3}{8} \(x^2+y^2+4xz+8z^2\)+\frac{2}{3}\e^u}, \nn
0\eql
        \frac{\d z}{\d u}+\frac{3z^2}{x}+\frac{3}{2}z+
            \frac{3}{8}\frac{x^2+y^2}{x}-
                \frac{2}{3}\frac{\e^u}{x} \label{hRiccati}
\ear
wherein we already have $x(u)$ and $y(u)$ from \re{gab} and \re{gob},
respectively.
\eq (\ref{hRiccati}) is a Riccati equation. On using the generalized
Riccati transformation \cite{noLie}
\be
    z=f\,\omega'(u)/\omega(u),
\ee
we obtain the second-order equation
\bearr
0 = f\frac{\omega''}{\omega}-f{u\frac{\omega'}{\omega}}^2
                        +f'\frac{\omega'}{\omega}
            +\frac{3}{x}f^2{\frac{\omega'}{\omega}}^2\nnn
\inch  +\frac{3}{8}\frac{x^2+y^2}{x}-\frac{2}{3}\frac{\e^u}{x}
\ear
and, setting $f={x}/{3}$
to remove the $\omega '^2$ terms, we eventually obtain
\bearr
0=\omega''+\(\frac{x'}{x}+\frac{3}{2}\)\omega'
        +\(\frac{9}{8}\frac{x^2+y^2}{x^2}
                    -2\frac{\e^u}{x}\)\omega,    \nnn
\ear
which is a linear second-order equation in $\omega$ as a function of $u$.
So we deduce $\dot \gg$ as a function of $\ga$,
\bear
\frac{\d u}{x}=\frac{\d w}{z} \ \iff\
I_4  \gg - \int {{\frac{\dot \gg}{\dot \ga}}\d a}.
\ear
Hence
\beq
t-t_0 = \int {\frac{\d a}{\sqrt{g(a)}}} \gb= \int{\frac{I_2}{\e^{3\ga}}\d t},
\eeq
so that this case is formally reduced to a quadrature and the solution of a
linear second-order differential equation.


\medskip\noi
{\bf Case 2:}
In this case the integrals have the same form. The only difference is that
$\gg$ depends on $\phi$. From the symmetry $\p _t$ we have the extra
integral
\be
        I_4=\e^{3\gg} {\dot \phi}
\ee
which can be solved to find
\be
        \phi=\int {I_4e ^{-3\gg} \d t}.
\ee

\medskip\noi
{\bf Case 3:}
We obtain exactly the same integrals as in the previous case.
We note that the second-order derivative of $\gg$ depends on
the constant potential.

\subsubsection{Bianchi Type V}

We consider here only the problem of finding the
Lie first integrals for Bianchi V in the first three cases, delaying the last
case to the next subsection. In Case 1
the integrals have the same expressions as those met
in the Bianchi I (with a different form of $f_1$ and
hence different $g_1$s).

\medskip\noi
{\bf Case 1:} The integrals are listed in Table 3
\begin{table}
\caption[75mm]{First integrals for Bianchi type V model in
the first case where there is no matter}
\begin{center}
\begin{tabular}{|c|c|}  \hline
Symmetries & Integrals \\  \hline
   \tall
$\p _t$ & $I_1=u^3 {\dot v}/{v}$  \\
    \tall
          & $I_2=u^3 {\dot w}/{w}$ \\
    \Tall \low
$v\p _v$ & \ \  $I_3=\Half{(\log u)}^3 \fracd{\dot u{}^2}{u} -f_1(u)$ \\
     \hline
\end{tabular}
\label{hIVcase1}
\end{center}
\end{table}
where
\bear
f_1(u) \eql  \frac{1}{4} \lb {(\log u)}^3+\frac{3(\log u)^2}{2}+
        \frac{3\log u}{2}
                    +\frac{3}{4} \rb  u ^{-2}  \nnn
  \nq    -  \frac{1}{48}\(3I_1 ^2 +I_2 ^2 \) \nnn
                                 \times
     \lb {(\log u)}^3 +\frac{1}{2}{(\log u)}^2
                + \frac{1}{6} \log u+\frac{1}{36} \rb u^{-6}.
\earn
Using the integral $I_3$, we find
\be
\dot u = {\sqrt {g_1(u)}}\ \iff\ t-t_0 = \int \frac{\d u}{\sqrt {g_1(u)}}
                \label{hintu}
\ee
where
\be
g_1(u)=\frac{2u^2 I_3}{{(\log u)}^3} + \frac{2u^2 f(u)}{{(\log u)}^3}.
\ee
From (\ref{hintu}) we can deduce $u(t)$. The variables $v$ and $w$
follow from the integrals $I_1$ and $I_2$, respectively. Hence we have
\be
v = \exp\int \frac{I_1}{u^3} \d t ,  \cm
w = \exp\int {\frac{I_2}{u^3}} \d t.
\ee
(This  solution uses only three
integrals since they are separable.)

\medskip\noi
{\bf Case 2:} The integrals are shown in Table 4
\begin{table}
\caption[75mm]{First integrals for Bianchi
type V model in the case where there is matter but no potential}
\begin{center}
\begin{tabular}{|c|c|}    \hline
Symmetries & Integrals\\  \hline
        \tall
$\p _t$ &     $I_1=u^3 {\dot v}/{v}$  \\
        \tall
          &     $I_2=u^3 {\dot w}/{w}$ \\
        \tall
          &     $I_3=u^3 \dot{\phi}$ \\
        \tall \low
$v \p _v$ & $I_4=\Half {(\log u)}^3 \dsp{\frac{\dot u}{u}} ^2 -f_2(u)$\\
     \hline
\end{tabular}
\end{center}
\end{table}
where
\bearr
f_2(u) = \frac{1}{4} \lb {(\log u)}^3+\frac{3{(\log u)}^2}{2}
+\frac{3\log u}{2}+\frac{3}{4} \rb  u ^{-2}
                    \nnn
    -\frac{1}{48}\(3I_1 ^2 +I_2 ^2 +2I_3 ^2 \)\nnn \quad
\times \lb
    {(\log u)}^3+\frac{1}{2}{(\log u)}^2
         + \frac{1}{6} \log u+\frac{1}{36} \rb u^{-6}.
\ear
Using the integral $I_4$, we see that
\be
 t-t_0= \int \frac{\d u}{\sqrt {g_2(u)}},\n
\ee
where
\be
g_2(u)=\frac{2u^2 I_3}{(\log u)^3} + \frac{2u^2 f_2(u)}{(\log u)}^3.\n
\ee
On the other hand, $I_1$ and $I_2$ give
\be
    v = \exp \int \frac{I_1}{u^3} \d t ,  \cm
    w = \exp \int {\frac{I_2}{u^3} \d t}.
\ee
Furthermore, using $I_3$, we obtain
\be
    \phi=\exp  \int \frac{I_3}{u^3} \d t.
\ee
We conclude that this case is reduced to quadratures.

\medskip\noi
{\bf Case 3:}
The integrals have exactly the same form as in the previous case with a new
function $f_3(u)$ replacing $f_2(u)$. We have
\bearr
f_3(u) =\frac{1}{4} \biggl[ {(\log u)}^3+\frac{3{(\log u)}^2 }{2}
    +\frac{3\log u}{2}+\frac{3}{4} \biggr] u^{-2}  \nnnv
        +\frac{1}{8}C{(\log u)}^4
     - \frac{1}{48}\(3I_1 ^2 +I_2 ^2 +2I_3 ^2 \)
                                     \biggl[(\log u)^3  \nnn  \cm\
 +\frac{1}{2}(\log u)^2
         + \frac{1}{6} \log u+\frac{1}{36} \biggr] u^{-6}.   \label{hf_3}
\ear

\subsection{Counteracting the symmetry breaking potential}

In this subsection we focus on Case 4. Note that, for instance,
the vector field $t\p _t$
is  a symmetry for Cases 1 and 2 but not for Cases 3 and 4
 and this is obviously due to the potential couplings.
In order to ``fix'' this problem  we consider an additional  term which
``kills'' any potential terms present in the $\gl$ and $\phi$ equations and
 becomes a symmetry for the other equations of the system.

Consider the vector field
\be
        v=t\p _t+a\p _{\phi},  \label{hnewsym}
\ee
where $a$ is a constant.
Note that this  is a symmetry for (\ref{hsys}b) and
(\ref{hsys}c) since the presence of the
additional term does not affect these two equations.
If we apply the new symmetry
to (\ref{hsys}a) and (\ref{hsys}d), we expect to find restrictions
on the  form  of the potential.

Applying the second extension of (\ref{hnewsym}) to (\ref{hsys}a),
 we find that the potential must satisfy the constraint
\be
    V''+\frac{2}{a}V'=0
\ee
and similarly from (\ref{hsys}d) we deduce that
\be
    V'+\frac{2}{a}V=0,
\ee
    giving immediately
\be
    V=K\e^{{-2\phi}/a}.
\ee
Recall that the potential terms are  responsible for reducing the number of
symmetries in Case 2. We now consider  the general  point symmetry of Case 2
for which there is no potential,
\bear
v \eql (A+Bt)\p _t+(C+Dt)\p _{\gl}\nnn
    +(E+F\phi+G\gb_2)\p _{\gb_1} +(H+I\phi-G\gb_1)\p _{\gb_2}   \nnn
    +\(J-\frac{3}{2}F\gb_1 -\frac{3}{2}I{\gb_2}\)\p _{\phi}
\ear
and apply its second extension
 to the set of equations where the potential
is unrestricted. \eq
(\ref{hsys}a) implies that $D=F=I=0$, which means that from the ten
initially possible symmetries we lose  three, the ones which  correspond
to the three vanishing constants. Also (\ref{hsys}d) gives
\bear
    V\eql K\e^{a\phi} ,  \\
    0\eql Ja+2B ,
\ear
so that one obtains  six independent constants as expected
(in Case 4 of this Type we found five independent symmetries). ????

A conclusion from this argument is that
when one applies the general form of the symmetry, one ends up with
the same symmetries as in Case 4 plus an additional one but with a
restricted potential.

Our method can indeed be used in more general situations.
More general  Bianchi cosmologies  have extra terms in the Euler-Lagrange
equations (\ref{hsys}) and one is forced to
introduce a symmetry with an extra term.
On considering the expression of the Ricci scalar in both classes and
following
the same reasoning as for Bianchi Type I, we are led to  the symmetry
\be
    v=t\p _t+a\p _{\phi}+\p_{\gl}. \label{hnew2sym}
\ee
If we take the second extension of (\ref{hnew2sym}),
we conclude that for both
classes $A$ and $B$ the potential is an exponential function of
$\phi$ and has exactly the same form as in Bianchi Type I.

\section{Conclusions}

The geometric reformulation of homogeneous cosmologies and the subsequent
applications discussed in the paper allow some more general conclusions to be
drawn regarding the dynamics of these cosmologies.
All models have in common the trivial
Noether symmetry $\p _t$ since all systems are autonomous.
Case 2 of most of the models   has the
common symmetry $\p_ \phi$. The symmetries that one  usually obtains
in Cases 2 and 3 are combinations of the symmetries in the other two cases.
Only Bianchi Type I in the case where the potential is constant gives a
symmetry which involves the constant potential. This cannot occur in
other models. Since Types VIII and IX are known to be the most complex, one
does expect this to be reflected in the calculation of their symmetries groups
and this was indeed the case as we
obtained only  the symmetries $\p _t$ and ${\p _\phi}$.

There are several directions in which one could extend research in this field.
Firstly, the symmetry group calculations performed here
can be extended to include other matter fields. Indeed, an interesting
project could be to consider the effects of incorporating a perfect fluid,
electromagnetic fields, \etc, on the number and nature of the symmetries
found here. A detailed analysis of the nature and number of contact
symmetries possibly present in homogeneous cosmologies would be also very
welcome.

The well-known dynamical aspects of Bianchi models must in some sense be
reflected in the symmetry groups discussed here. How do our symmetries behave
as these models expand? One does  expect that neither their number nor their
nature  remains intact asymptotically towards or away from singularities.
Besides, some models are known to isotropise. Do their {\em symmetries}
evolve towards precisely those of the associated FRW models? We conjecture
that they do, but this is a problem for the future.

\Acknow
{S.C. thank the Research Commission of the University  the Aegean for its
continuing support. P.G.L.L. thanks Prof. G.P. Flessas of the Department
of Mathematics at the University of the Aegean for his kind hospitality and
the Foundation for Research and Development of South Africa and the
University of Natal for their continuing support.    The work of H.P. was
supported by a grant from the General Secretariat for Science and Technology
which is gratefully acknowledged.}


\small
\begin{thebibliography}{99}

\bibitem{o1}
      P.J.  Olver, ``Applications of Lie Groups to Differential
      Equations'', 2nd Edition, Springer, New York, 1993.

\bibitem{o2}
     P.J. Olver, ``Equivalence, Invariants and Symmetry'', CUP,
     Cambridge, 1995.

\bibitem{baco}
     J.D. Barrow and S. Cotsakis, {\it Phys. Lett.\/} {\bf B214}, 515 (1988).

\bibitem{italians}
     S. Capozzielo, G. Marmo, C. Rubano and P. Scudellaro,
     {\it Int. J. mod. Phys.\/} {\bf D6}, 491 (1997).

\bibitem{wym}
     W. Hereman, {\it Euromath. Bull.\/} {\bf 1}, 45 (1994).

\bibitem{lie}
     S. Lie, ``Theorie der Transformationsgruppen'', Vols. I, II, III,
     Chelsea, New York, 1970.

\bibitem{nl1}
     B. Abraham-Shrauner and P.G.L. Leach, {\it in:}
    ``Exploiting Symmetry in
     Applied and Numerical Analysis'', eds. E. Algower, K. Georg and R.
    Miranda, Lectures in Applied Mathematics {\bf 29}, AMS, Providence,
    1993.

\bibitem{nl2}
     K.S. Govinder and P.G.L. Leach, {\it J. Phys. A: Math. Gen.\/} {\bf 30},
     2055 (1997).

\bibitem{noe}
     E. Noether,  {\it Kgl. Ges. d. Wiss. Nachrichten Math-phys. Kl.\/}
     {\bf Heft 2}, 235 (1918).

\bibitem{p81}
     P.G.L. Leach, {\it J. Austr. Math. Soc.\/} {\bf 23}, 173 (1981).

\bibitem{maccallum}
     M.A.H. MacCallum, {\it in:} ``General Relativity: An Einstein
     Centenary Survey'', eds. S.W. Hawking and W. Israel,
    CUP, Cambridge, 1979.

\bibitem{head}
     A.K. Head, {\it Comp. Phys. Comm.\/} {\bf 77}, 241 (1993).

\bibitem{noLie}
     B. Abraham--Shrauner, K.S. Govinder and P.G.L. Leach, {\it Phys.
     Lett.\/} {\bf B203}, 169 (1995).

\end{thebibliography}
\end{document}